\documentclass[10pt, Journal]{IEEEtran}
\usepackage{setspace}
\usepackage{amsfonts}
\usepackage{dsfont}
\usepackage{cite}
\usepackage{graphicx}
\usepackage{amssymb}
\usepackage{stfloats}
\usepackage{booktabs}
\usepackage{verbatim}

\usepackage{paralist}
\usepackage{amsmath}
\usepackage{amssymb}
\usepackage{lpic}
\usepackage{color}
\usepackage{bm}
\usepackage[ruled,boxed,linesnumbered,vlined]{algorithm2e}
\usepackage{url}
\usepackage{balance}
\usepackage{amsthm}
\usepackage{amsfonts}

\usepackage{cases}
\usepackage{epstopdf}
 \usepackage{float}
 \usepackage{lipsum}
 
 \usepackage{subfig}
\usepackage{caption}
\usepackage{graphicx}
\usepackage{float} 
\usepackage[justification=centering]{caption}

\usepackage{amsmath}

\def\BibTeX{{\rm B\kern-.05em{\sc i\kern-.025em b}\kern-.08em
    T\kern-.1667em\lower.7ex\hbox{E}\kern-.125emX}}
\begin{document}
\title{Spectrum-aware Multi-hop Task Routing in Vehicle-assisted Collaborative Edge Computing\footnotemark}


\author{
\IEEEauthorblockN{Yiqin~Deng,~\IEEEmembership{Member,~IEEE,} Haixia~Zhang,~\IEEEmembership{Senior Member,~IEEE,} Xianhao~Chen,~\IEEEmembership{Member,~IEEE,}  and~Yuguang~Fang,~\IEEEmembership{Fellow,~IEEE}}
\thanks{This work was supported in part by the Project of International Cooperation and Exchanges NSFC under Grant No. 61860206005 and in part by the Joint Funds of the NSFC under Grant No. U22A2003. (\emph{Corresponding author: Haixia Zhang})}
\thanks{Yiqin Deng and Haixia Zhang are with the Shandong Key Laboratory of Wireless Communication Technologies, Jinan, Shandong, 250061, China, and also with
the School of Control Science and Engineering, Shandong University, Jinan,
Shandong, 250061, China (email: yiqin.deng@email.sdu.edu.cn; haixia.zhang@sdu.edu.cn). }
\thanks{Xianhao Chen is with the Department of Electrical and Electronic Engineering, University of Hong Kong, Hong Kong, China (email: xchen@eee.hku.hk).}
\thanks{Yuguang Fang is with the Department of Computer Science, City University of Hong Kong, Hong Kong, China (email: my.fang@cityu.edu.hk).}
}
\maketitle

\begin{abstract}
Multi-access edge computing (MEC) is a promising technology to enhance the quality of service, particularly for low-latency services, by enabling computing offloading to edge servers (ESs) in close proximity. To avoid network congestion, collaborative edge computing has become an emerging paradigm to enable different ESs to collaboratively share their data and computation resources. However, most papers in collaborative edge computing only allow one-hop offloading, which may limit computing resource sharing due to either poor channel conditions or computing workload at ESs one-hop away. By allowing ESs multi-hop away to also share the computing workload, a multi-hop MEC enables more ESs to share their computing resources. Inspired by this observation, in this paper, we propose to leverage  omnipresent vehicles in a city to form a data transportation network for task delivery in a multi-hop fashion. Here, we propose a general multi-hop task offloading framework for vehicle-assisted MEC where tasks from users can be offloaded to powerful ESs via potentially multi-hop transmissions. Under the proposed framework, we develop a reinforcement learning based task offloading approach to address the curse of dimensionality problem due to vehicular mobility and channel variability, with the goal to maximize the aggregated service throughput under constraints on end-to-end latency,  spectrum, and computing resources. Numerical results demonstrate that the proposed algorithm achieves excellent performance with low complexity and outperforms existing benchmark schemes.
\end{abstract}

\begin{IEEEkeywords}
Computation offloading, Collaborative edge computing, Vehicular networks, Multi-hop service request routing, Deep reinforcement learning (DRL).
\end{IEEEkeywords}

\section{Introduction}
\label{sec:introduction}
Multi-access edge computing (MEC) has been identified as a promising architecture for computing services that aims to provide real-time or low latency services to end-users located in close proximity~\cite{deng2022actions,wang2020convergence}. One of the primary techniques utilized in MEC is computation offloading, which enables computing tasks to be processed locally or offloaded to an edge server (ES) based on the availability of local computing resources and transmission conditions~\cite{lin2019computation}. This approach is beneficial for mobile devices (MDs) that are typically limited in terms of their computing capability, storage capacity, and battery power~\cite{zhou2019energy,pliatsios2022joint}. Moreover, through effective computation offloading in MEC, end-to-end (e2e) latency for emerging capability-demanding or latency-sensitive  applications can be drastically reduced, ultimately providing high quality-of-services to end users~\cite{deng2019parallel,wang2022joagt}. 


To optimize resource utilization and efficiency, extensive research efforts have been dedicated to addressing computation offloading in MEC~\cite{cao2018revisiting, poularakis2020service, deng2021throughput,li2020learning, li2020deep,sahni2021multi}. Previous research efforts have primarily focused on computation offloading and resource optimization that directly associate users with ESs within a user's communication range (i.e., one-hop away ESs), considering computing resources and/or communication resource optimization at a single ES~\cite{cao2018revisiting, poularakis2020service, deng2021throughput}. However, this approach may fall short in practice due to the lack of coordination among ESs, which are required for better load balancing~\cite{lin2023efficient}. Despite research efforts attempting to enable resource-constrained ESs to collaborate in processing computation-intensive tasks to achieve workload balancing, the current literature is primarily concerned with one-hop offloading between MDs and ESs~\cite{li2020learning, li2020deep,sahni2021multi}. Such an approach incurs an implicit assumption, whereby communication resources available at the ESs one-hop away are sufficient for uploading complex computing tasks. This suppositional approach may not always be effective in reality. Specifically, the single-hop offloading approaches may not work well under resource-constrained scenarios. For example, when considering MEC-enabled surveillance video analytics for public safety applications in smart cities~\cite{pang2019spath}, where a large volume of high-resolution videos should be transported from street cameras to distributed MEC servers for processing, task uploading and/or computing may fail when the spectrum/computing resources  are insufficient to serve the service demands at the spot. To enhance resource utilization, a data transportation network is needed for task delivery from end users to appropriate edge servers with available computing and spectrum resources via potentially multi-hop delivery~\cite{ding2021probabilistic}. Therefore, to optimize resource utilization and efficiency, future research efforts in MEC systems ought to explore better coordination and collaboration between MDs and ESs to address computation offloading issues, which will lead to improved throughput performance, better workload balancing, and ultimately, better resource utilization~\cite{tang2021survey}. 

Aiming at small computing latency while avoiding network congestion, Dai~\emph{et al.}~\cite{dai2023task} recently proposed a cooperative offloading framework in device-to-device (D2D)-assisted MEC networks, where both ESs and idle MDs enable offloading services for computing-intensive industrial tasks. Here, each user delivers offloading service for at most one neighbor MD to avoid queueing latency as the communication coverage by D2D links is small. Chukhno~\emph{et al.}~\cite{chukhno20235g} claimed that reliability in public safety services can be achieved via multi-hop relaying, which is considered to be one of the key technologies facilitating enhanced system performance in future 5G+ systems. For example, it allows establishing direct connections between devices in scenarios outside the coverage area, thus ensuring first responders with the connectivity they need, especially in hazardous situations. As discussed in~\cite{garcia2021tutorial}, the Third Generation Partnership Project (3GPP) has already identified new study and work items for New
Radio (NR) Vehicle-to-Everything (V2X) side-link (SL) communication within Release 17, among which the concept of MD relaying has been proposed to extend the coverage range. Different from utilizing a single relay, which is referred to in 3GPP as a single-hop NR SL-based relay, forward compatibility for multi-hop relay support in a future release will be taken into account~\cite{garcia2021tutorial}. These standardization progresses demonstrate the importance and feasibility of relaying in vehicular networks, which can naturally be leveraged for computing task offloading.  

Motivated by the performance enhancement brought by such multi-hop D2D transmissions, in our prior work~\cite{deng2021leverage}, we have explored vehicle-assisted multi-hop transmissions to balance the computing workload at ESs under a simple scenario with one MD and multiple ESs. In this paper, we propose to employ vehicles ubiquitously available in a city to form a data transportation network, which could facilitate multi-hop task offloading between a user and the associated ES. Due to the omnipresence of vehicles, this approach is economically sound because no additional fixed relays are needed. Moreover, thanks to the short device-vehicle and inter-vehicle distances, MDs and vehicles can employ short-range multi-hop transmissions with low transmit power, thereby causing less interference and improving network-wide spectrum reuse~\cite{noor20226g}. When extending to a more general case, new challenges will arise. One critical issue in multi-hop offloading is to make a trade-off between the communication overhead and the computing capability to satisfy quality-of-service (QoS) requirements~\cite{chen2022federated,chen2022end}, which will increase the complexity of the task offloading problem. Such an issue is further complicated by the dynamic nature of network topology due to vehicular mobility~\cite{liu2021livemap,qin2022learning}.

To fill in this gap, this paper investigates the problem of spectrum-aware task offloading in vehicle-assisted multi-hop edge computing. The challenges are fourfold. First, the vehicular network environment is highly complicated and dynamic, which can hardly be captured by an accurate and mathematically solvable model. Thus, traditional task offloading methods are not suitable under this scenario. Second, dynamic task routing decisions are jointly made with task-server assignments, which is more challenging than traditional routing with predetermined source and destination nodes. Third, the network-wide tradeoff between the communication and computing workloads further complicates the task offloading problem. At last, typical solutions based on queueing theory may not work well for the situation involving multi-hop routing and e2e QoS guarantees because a few strong assumptions (e.g., task arrivals at every source node and intermediate node follow a Poisson process) underpinning the analytical results may not hold and many problems in multi-point to multi-point queueing networks still remain open~\cite{xu2018experience}.

To tackle the above challenges, we first propose a general multi-hop task offloading framework for a vehicle-assisted MEC network, where different e2e paths can be simultaneously established between users and remote ESs via multi-hop transmissions by utilizing different groups of relay vehicles and the destination ES. This gives rise to a new task routing design problem in which the selected vehicles, the target ES, and their routing paths for different users need to be jointly optimized to balance the workloads in terms of both communication and computing to maximize the processed task size over the whole system. Although the system capacity can be enhanced by coordinating the network-wide resources, it is hard to guarantee service reliability with the uncertainty of vehicular trajectories. To deal with this issue, we resort to the multi-agent deep deterministic policy gradient (MADDPG) method, a deep reinforcement learning (DRL), which is capable of addressing issues with high dimensional states and huge action spaces~\cite{li2022mec}. To this end, we present a novel and highly effective MADDPG-based task offloading approach for a multi-hop MEC by learning network dynamics. Note that our approach is not restricted to vehicle-aided MEC
and it can be easily extended to other multi-hop MEC systems facing similar challenges. 

Our main contributions can be summarized as follows.
\begin{itemize}
    \item We are the first to present the framework of multi-hop task offloading by coordinating the resources in a multi-hop multi-edge multi-user MEC system. Under the proposed framework, we formulate a throughput maximization problem subject to both communication and computing resource constraints and e2e latency requirements.
    
    \item To solve the original optimization problem, we reconstruct a Markov decision process (MDP) based formulation for the task offloading decision-making under uncertainty of network topology.
    
    \item We resort to a model-free DRL method, i.e., MADDPG, to find an effective task offloading solution under dynamic spectrum and computing resource constraints by learning the undetermined model via interactions with a vehicle-assisted MEC environment. 
\end{itemize}

The remainder of this paper is organized as follows. In Section II, we present the related works. Section III describes the system model and problem formulation. In Section IV and V, we present the preliminaries for MADDPG and the MADDPG-based task offloading scheme, respectively. Section VI presents simulation results, and Section VI concludes this paper.  

\section{Related work}
\label{sec:related work}
Most existing works on computation offloading in MEC focus on single-hop offloading from MDs to ESs. They can be roughly divided into two categories according to whether the cooperation between ESs is involved: \textit{resource optimization for an MEC with a single ES}~\cite{cao2018revisiting, poularakis2020service, deng2021throughput} or \textit{cooperative MEC over multiple ESs}~\cite{li2020learning, li2020deep}. In this section, we will first review the research status according to the above two categories, and then survey the related works on multi-hop task offloading from MDs to ESs.

\subsection{Resource optimization for an MEC with a single ES}
Cao~\emph{et al.}~\cite{cao2018revisiting} proposed computation partitioning, dispatching, and scheduling algorithms for 5G-based edge computing systems, under the assumption that there is plenty of spectrum bandwidth, to support the data transmissions between MDs and an ES, which could parallelize computing tasks and fully utilize the computing resources at both the ES and MDs. Based on the observations that a considerable amount of data should be pre-stored and asymmetric spectrum bandwidth is required for uplink and downlink transmissions to support many emerging services (e.g., Augmented Reality (AR) services) at ESs, Poularakis~\emph{et al.}~\cite{poularakis2020service} studied
the joint optimization of service placement and computation offloading for MEC networks with storage, computation, and communication constraints. While the optimization problem here is probably the most general one to minimize the computing workload offloaded to the centralized cloud under the above system consideration, they did not consider queueing at ESs, which is commonly encountered in practical systems. In~\cite{deng2021throughput}, Deng~\emph{et al.} proposed a scheme to maximize the task completion ratio (throughput) in MEC under e2e latency constraints by using a tandem queue model to characterize the joint resource allocation of communications and computing. They also considered the stochasticity of involved processes, e.g., task arrivals, random channels, and varying computing power. However, this paper only focuses on a single MD and single ES scenario. 

It is also observed that all these works only study the scenarios that the computing tasks can directly be transmitted to the destination ES within the MD’s communication range (i.e., one-hop) at one single ES without considering the cooperation among ESs.

\subsection{Cooperative MEC over multiple ESs}
By exploiting cooperation among ESs, tasks that arrive at one ES can be either processed locally or partially/fully offloaded to powerful ESs via backbone or backhaul links to enhance the quality of experience (QoE). In~\cite{li2020learning}, Li~\emph{et al.} proposed an online cooperative offloading mechanism to optimize the decision of task admission and scheduling among ESs with the objective to minimize the long-term system cost by considering full offloading (i.e., binary offloading).
In~\cite{li2020deep}, Li~\emph{et al.} extended the cooperative computing framework in MEC to vehicular networks by considering challenges in computing result
delivery due to the uncertainty of vehicular mobility. To
address the complexity resulting from the dynamic network topologies in MEC-enabled vehicular networks, they proposed a location-aware offloading and computing strategy to coordinate ESs with partial offloading (i.e., computing at multiple ESs in parallel). However, they still assume that backbone/backhaul links have plenty of bandwidth, and hence will not pose any constraints on communications between ESs.

It is also observed that all these works enable resource-constrained ESs to help each other in processing computation-intensive tasks, thereby enhancing computing workload balancing and resource utilization in MEC systems.  

\subsection{Multi-hop task offloading between MDs and ESs}
The aforementioned research works generally make an implicit assumption that MDs can only offload tasks to ES one-hop away, which significantly restricts the solution space and limits resource sharing. For example, when a nearby server is overwhelmed with its computing, it is natural to offload a MD’s task to other servers potentially unreachable by one-hop communications, or when too many MDs at one-hop away ESs are excessive, there is no spectrum used to offload data to one-hop away ESs, while there may exist multi-hop path connecting to multi-hop away ESs. In either case, multi-hop offloading may be leveraged to increase resource sharing and load balancing.

As far as we know,~\cite{hui2020reservation} and~\cite{deng2021leverage} are probably the most related works tackling vehicle-assisted multi-hop task offloading as done in this paper. In~\cite{hui2020reservation}, Hui~\emph{et al.} designed a request relay mechanism for MEC-enabled vehicular networks to reduce the cost of the relay service by taking the dynamic traffic conditions and the reputation of vehicles into consideration. However, they merely considered the limited transmission ranges of vehicles and ESs while ignoring resource constraints and QoS requirements. In~\cite{deng2021leverage}, Deng~\emph{et al.} proposed a load-balanced relay mechanism for MEC in which the relay vehicle and destination ES are jointly determined according to the queueing status at an MD and traffic status, significantly enhancing the system performance. Nevertheless, Their work just considered a simple case with a single MD where the complicated task routing between multiple MDs and destinations is not involved. Different from these works, this paper intends to employ vehicles as relays for computing task delivery by taking advantage of the mobility and spectrum opportunities in vehicular environments. To deal with the ``curse of dimensionality'' arising from large-scale vehicular networks, we use DRL to find the multi-hop task routing paths. 


\section{System Model and Problem Formulation\label{system_model}}
This section describes the proposed multi-hop task offloading framework in vehicle-assisted MEC. 

\subsection{Multi-hop task offloading framework in vehicle-assisted MEC}
\begin{figure}[t]
\includegraphics[width=0.48\textwidth]{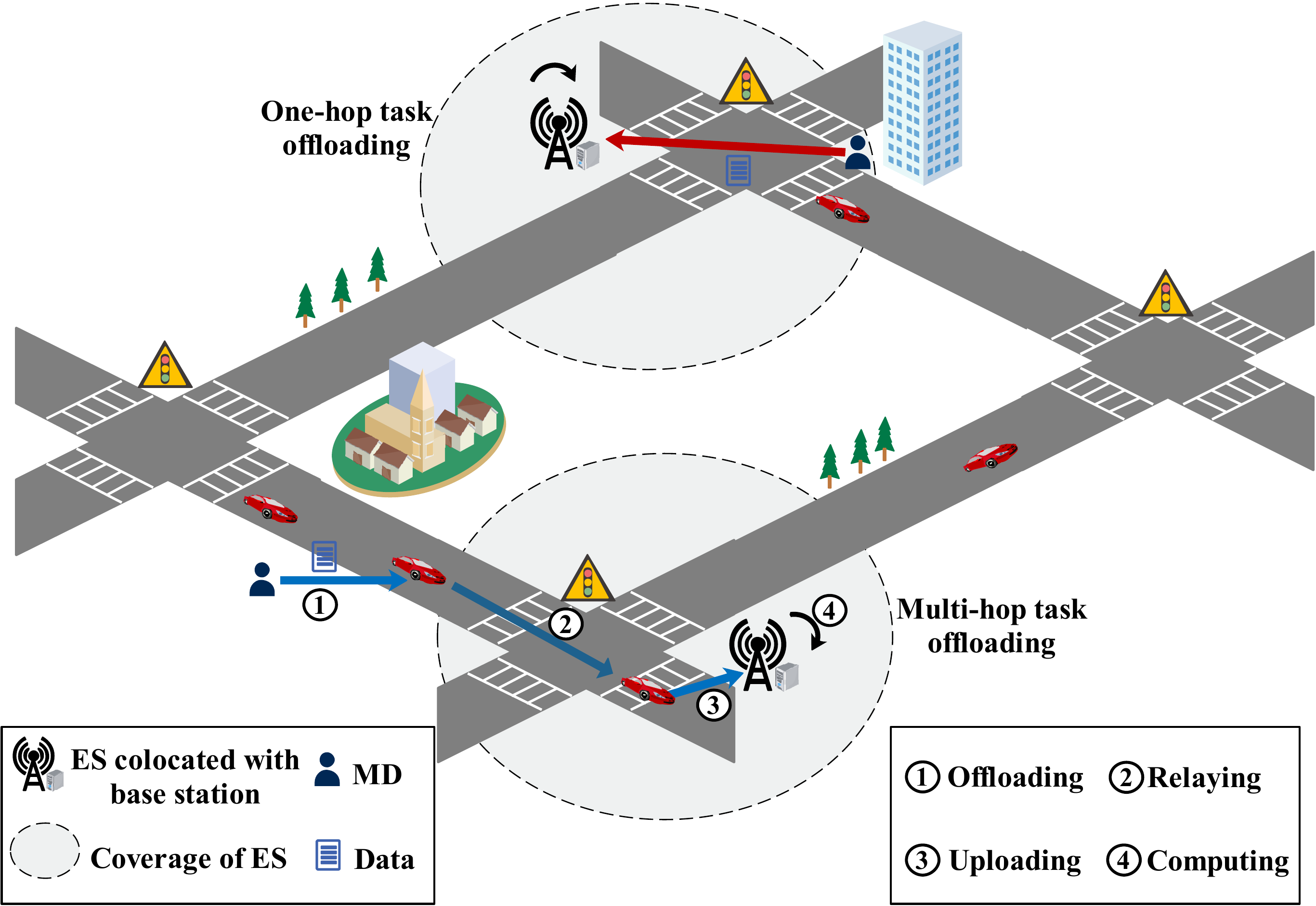}
\centering
\caption{A multi-hop task offloading framework for a vehicle-aided MEC system.}
\label{model}
 \end{figure}

As shown in Fig.~\ref{model}, we consider an MEC system with multiple MDs and ESs, where $N$ distributed vehicles are deployed to assist in the communication from MDs to $J$ remote ESs. Without loss of generality, we assume that i) MDs in this system can be either pedestrians on the roadside or passengers in the vehicle, and ii) tasks from MDs can either be offloaded to the ES within their communication range or offloaded to a remote ES via multi-hop transmission path assisted by vehicles. We consider a time-slotted system $t\in \mathcal{T}=\{0,1,2,\cdot\cdot\cdot,T\}$. Suppose that each MD is associated with at most one ES at any given time while each ES can serve multiple MDs via proper user scheduling. As such, we focus on the design of a routing path for tasks generated from a set of MDs in one given time slot. For convenience, we denote the sets of MDs, vehicles, and ESs as $\mathcal{I}=\{1,2,\cdot\cdot\cdot,I\}$, $\mathcal{N}=\{1,2,\cdot\cdot\cdot,N\}$, and $\mathcal{J}=\{1,2,\cdot\cdot\cdot,J\}$, respectively.

MD $i$ may fail to access the service of an MEC server within its deadline $D_i$ because i) computing workload at the surrounding ESs is excessive or ii) the ES is out of its communication range or iii) the channel condition between MD $i$ and the ES is poor or transmission channel between MD $i$ and its surrounding ESs are excessively busy. Thus, it is possible that MD $i$ seeks help from vehicles on the road to relay its data to an appropriate ES multi-hop away, improving the system capacity. We assume that a global controller has global knowledge of the network dynamics and makes offloading decisions for all users in a centralized manner. For example, we could take software-defined networking (SDN) design approach to implement our proposed MEC systems. To conclude, a multi-hop MEC service session for a task includes the following four steps while we ignore the procedure of result returning since the size of results in many practical applications (e.g., object detection results) is relatively small.

1) \textit{Offloading}: when a computing task is generated at MD, it selects a relay vehicle within its communication range and offloads the computing data
of the task to the vehicle.

2) \textit{Relaying}: after a vehicle receives the computing data from the MD, it transfers data across vehicles on the road by choosing an appropriate route to the destination ES.

3) \textit{Uploading}: when the relay vehicle arrives in the communication range of the destination ES, it uploads the carried data to the ES. 

4) \textit{Computing}: after the computing data is fully offloaded, the destination ES can process the computing task and send back the result to the MD after finishing the computing.


Let $p_i \in \mathcal{P}$ denote the route from MD $i$ to a destination ES, where $\mathcal{P}$ is the set of feasible routes and it consists of none (i.e., one-hop transmission from MD $i$ to the ES) or multiple vehicles (i.e., multi-hop routing) and one ES. Let $a_{i,p_i}(t)=\{0,1\}$ denote the offloading decision for MD $i \in \mathcal{I}$ at time slot $t$. Thus, we have
\begin{equation}\label{1}
\sum \limits_{p_i\in\mathcal{P}}a_{i,p_i}(t) \leq 1, \forall i,t.
\end{equation}

According to the propagation model in 3GPP standards~\cite{li2020deep}, the path loss between a transmitter and a receiver with distance $d$ (km) can be computed as:
\begin{equation}\label{2}
\begin{split}
\Psi(d)=&40\left(1-4\times10^{-3}H\right)\log_{10}d - 18\log_{10}H\\&+21\log_{10}f+80(dB),
\end{split}
\end{equation}
where $H$ and $f$ are the antenna height in meter and the carrier frequency in MHz, respectively. The distance between node $a$ and $b$ is denoted as $D_{a,b}$. Thus, from the Shannon capacity theorem, the data rate between node $a$ and $b$ can be expressed as:
\begin{equation}\label{3}
R_{a,b}(t)=B\log_2\left(1+\frac{P*10^{-\Psi(D_{a,b})/10}}{\sigma^2}\right),
\end{equation}
where $\sigma^2$ denotes the power of the Gaussian noise in the channel (e.g., the user-to-vehicle channel, the vehicle-to-vehicle (V2V) channel, or the vehicle-to-infrastructure channel), $P$ represents the node's transmit power, and $B$ represents the spectrum bandwidth used by the MD. 

Suppose the size of the data generated by MD $i$ in time slot $t$ is $W_{i}(t)$, we have the latency when delivering the data from MD $i$ to the first relay vehicle $n^{first}$:
\begin{equation}\label{4}
L^{trans}_{i,n^{first}}(t)=\frac{a_{i,p}W_{i}(t)}{R_{i,n^{first}}(t)}.
\end{equation}

Moreover, the latency when relaying the data between two adjacent vehicles in the route $p_i$, e.g., vehicles $n$ and $n'$, is:
\begin{equation}\label{6}
L^{trans}_{n,n'}(t)=\frac{a_{i,p_i}W_{i}(t)}{R_{n,n'}(t)}.
\end{equation}

Accordingly, the latency when uploading the data from the last relay vehicle $n^{last}$ to the destination ES $j$ is:
\begin{equation}\label{8}
L^{trans}_{n^{last},j}(t)=\frac{a_{i,p}W_{i}(t)}{R_{n^{last},j}(t)}.
\end{equation}

Let $\mathcal{P}^v$ denote the set of adjacent vehicle pairs, e.g., vehicle pair $(n,n')$, in the route $p_i$. Thus, the latency when forwarding the data from MD $i$ to the destination ES $j$ can be expressed as:
\begin{equation}\label{9}
L^{trans}_{i,p_i}(t)=L^{trans}_{i,n^{first}}(t)+\sum\limits_{(n,n')\in \mathcal{P}^v} L^{trans}_{n,n'}(t)+L^{trans}_{n^{last},j}(t).
\end{equation}

After the task from MD $i$ is delivered and other tasks in the queue has been completed, the task can be processed by the dedicated ES. The latency when accomplishing task $i$ generated in time slot $t$ on ES $j$ can be formulated as
\begin{equation}\label{10}
L^{comp}_{i,p_i}(t)=\frac{\kappa a_{i,p_i}(t)W_{i}(t)}{C_j},
\end{equation}
where $\kappa$ is the computation cycle per bit data.

Besides, the queueing latency of task from MD $i$ at ES $j$ is defined as the latency for finishing the uncompleted tasks offloaded in previous time slots $\{1,2,\cdot\cdot\cdot,t-1\}$, which can be formulated as follows:
\begin{equation}\label{11}
L^{queue}_{i,p_i}(t)=\max \left\{\sum \limits_{i'\in \mathcal{I}\slash i}L^{comp}_{i',p_i}(t-1)-\epsilon, 0\right\},
\end{equation}
where $\epsilon$ is the duration of a time slot.

Given the transmission latency, computing latency, and queueing latency, the e2e service latency for MD $i$ can be formulated as follows:
\begin{equation}\label{12}
L_{i,p_i}(t)=L^{trans}_{i,p_i}(t)+L^{comp}_{i,p_i}(t)+L^{queue}_{i,p_i}(t).
\end{equation}

\subsection{Problem formulation}
Our objective is to find a task routing policy $ \boldsymbol{\alpha}$ with the goal to maximize the aggregated throughput for the vehicle-assisted multi-hop MEC system while guaranteeing the end-to-end latency requirements from MDs. The optimization objective is thus formulated as

\begin{equation}
 \max_{\alpha}\quad \sum \limits_{t\in\mathcal{T}}\sum\limits_{i\in\mathcal{I}}\sum \limits_{p_i\in\mathcal{P}} a_{i,p_i}(t)W_i(t) \cdot {1}_{\{L_{i,p_i}(t) \leq D_i\}}\label{ob}
\end{equation}
where $\mathds{1}_{\{L_{i,p_i}(t) \leq D_i\}}$ is the indicator function whose value takes 1 when the e2e service latency requirement of MD $i$ is satisfied, or 0 otherwise. Note that $L_{i,p_i}(t)$ is calculated after the task generated by user $i$ at time slot $t$ is accomplished in the current or the future time slot. Moreover, the objective function in~\eqref{ob} represents the total size of the tasks completed with latency requirements during the considered time duration $T$. With the optimization objective in~\eqref{ob}, we have to take into account communication and computing resource constraints. 

To solve problem~\eqref{ob}, we are facing three nontrivial challenges. First, it is hard to express the e2e latency in a closed form. As we mentioned in Section I, the queueing theory as a typical tool to address the e2e latency may not fit the problem of multi-hop task offloading since the task arrival assumption at every node fails to hold for the classical queueing models with multi-point to multi-point to have close form solutions. Second, it is infeasible to directly solve problem~\eqref{ob} by the traditional optimization method since it is a mixed-integer non-linear optimization problem. Third, even though we can model the system as a Markov decision process, it is hard to overcome the curse of dimensionality in terms of both state space and action space. Taking the action variable as an example, i.e., $a_{i,p_i}(t)$, there exists $I \times N \times J $ 
 decisions to make in one time slot. Moreover, it is not efficient to use the traditional queueing theory to handle the problem with multiple outputs while the task offloading policy for each MD is given simultaneously in problem~\eqref{ob}.

Based on the above analysis, we resort to the deep deterministic policy gradient (DDPG) method to address these challenging issues since DDPG is not only good at solving optimization problems with large action spaces, but also has a good convergence performance as demonstrated in ~\cite{lillicrap2016continuous}.

\section{Preliminaries for Deep Deterministic Policy Gradient}
\subsection{MDP-based Task Offloading Model}
To solve problem~\eqref{ob} with DDPG, we first model it as an Markov decision process (MDP) $(\mathcal{S},\mathcal{A},P,R)$, where $\mathcal{S}$ and $\mathcal{A}$ are the sets of system states and actions, respectively, and $P$ and $R$ are the functions of state transition and reward, respectively. The specific definitions are given below.


 \emph{State space:} The design of state space is to reflect the status of the considered system completely and informatively. Therefore, we build the state space $\mathcal{S}$ consisting of \textit{vehicle status}, \textit{server status}, and \textit{system workload}. \textit{Vehicle status} provides the information of the feasible relay vehicles and the channel states among vehicles. \textit{Server status} includes the computing capability of ESs and the available bandwidth. \textit{System workload} provides the information of the amount of input data from MDs and the number of queueing tasks at ESs.



 \emph{Action space:} Based on the observed state, the actions can be chosen from the feasible action space $\mathcal{A}$ in each time slot whose element represents the routing path for each MD.

\emph{Transition probability:} Transition probability in MDP represents the probability that the system state moves from the current state $s$ to the next state $s^\prime$ when action $a$ is taken, i.e., $P^a_{ss^\prime}=\mathbb{P}\{s^\prime|(s,a)\}$.

\emph{Reward:} In an MDP, the reward is related to both state and action. When an action, e.g., a task scheduling policy, is selected under the current state, the corresponding reward will be received from the system, i.e., $R^a_{s}=\mathbb{E}\{R|(s,a)\}$. In the considered problem, the reward function can be set according to the objective function~\eqref{ob}.

For the MDP, $\pi(s,a):\mathcal{S}\times \mathcal{A}\rightarrow [0,1]$ is set to a policy that  gives the probability of taking action $a$ when in the state $s$. To obtain the expected long-term discounted reward, the value function $Q$ of state $s$ by taking policy $\pi$ is
\begin{equation}
Q(s,\pi)=\mathbb{E}\left[\sum \limits_{t\in \mathcal{T}} \gamma^t R^a_{s}(t) \right],
\end{equation}
where $\gamma \in [0,1)$ is a discounting factor. By maximizing the value function across different states, we can obtain the optimal task scheduling policy $\pi^\ast$:




\begin{equation}\label{MDP_ob}
\pi^\ast(s,a)=\arg \max\limits \sum \limits_{s'}\mathbb{P}(s'|(s,a))\left[R(s,a)+\gamma Q(s',\pi^\ast)\right].
\end{equation}


\subsection{Deep Deterministic Policy Gradient}
For the optimization in~\eqref{MDP_ob}, the traditional dynamic programming cannot find the optimal policy as we have no knowledge about the transition probability $P$ in the considered system. Therefore, we resort to model-free reinforcement learning, i.e., deep deterministic policy gradient (DDPG)~\cite{lillicrap2016continuous}, to learn the model via the interactions between agents and the environment.

In DDPG, there are a total of four networks: the Actor, the Critic, and the corresponding target networks for the Actor and Critic, respectively. The target networks can be regarded as time-delayed copies of their original networks that slowly track the learned networks, which will significantly enhance the stability of learning. The specific functions of these four neural networks are as follows.

1) \emph{Actor network:} The Actor network is in charge of the iterative update of policy network parameters and the direct maps from the current state to the current action. In this way, it interacts with the vehicle-assisted multi-hop MEC environment to generate the next state and reward.

2) \emph{Actor target network:} The Actor target network outputs the next optimal action according to the next state sampled in the experience replay. The network parameters in the Actor target network are periodically copied from the Actor network.

3) \emph{Critic network:} The Critic network is responsible for the iterative update of the parameters in the value network and calculating the current $Q$ value.

4) \emph{Critic target network:} The Critic target network calculates $Q^\prime$ value according to the next state-action. The network parameters in the Critic target network are periodically copied from the Critic network.

The above two target networks have ``soft''-updates based on main networks, i.e., the target networks only update a small part based on the current network, to improve the stability of learning. That is, 
\begin{equation}
\theta^\prime \leftarrow \tau\theta+(1-\tau)\theta^\prime,
\end{equation}

\begin{equation}
w^\prime \leftarrow \tau w+(1-\tau)w^\prime,
\end{equation}
where $0\textless \tau \ll 1$ is the update frequency for the parameters in \emph{actor target network} ($\theta$) and \emph{critic target network} ($w$).

To improve the exploration capability and thus avoid getting stuck in a local optimum, DDPG typically adds noise ($\mathcal{N}_t$) to the action ($\mathcal{\pi}_\theta(s)$) produced by the \emph{actor network} to get a new action, i.e.,
\begin{equation}
a=\mathcal{\pi}_\theta(s)+\mathcal{N}_t.
\end{equation}

The loss functions for the \emph{critic network} and the \emph{actor network} are respectively defined as
\begin{equation}\label{loss_Critic}
L(w)=\frac{1}{m}\sum \limits_{z=1}^m (y^z-Q(\phi(S^z),A^z,w))^2,
\end{equation}
and
\begin{equation}\label{loss_Actor}
L(\theta)=-\frac{1}{m}\sum \limits_{z=1}^m Q(s,a,\theta),\;\; z=1,2,\cdot\cdot\cdot,m,
\end{equation}
where $m$ is the number of samples (including eigenvector of state $\phi(S^z)$, and action $A^z$) from Replay Buffer $\mathcal{D}$, $y^z$ is the target value of $Q$, 

\section{MADDPG-based Task Routing in Vehicle-assisted MEC}
In this section, we elaborate how to leverage DDPG to solve our task routing problem. Here we exploit Multi-Agent Deep Deterministic Policy Gradient (MADDPG) to cope with the curse of dimensionality in the multi-hop task offloading optimization problem. To tackle the problem efficiently, the problem is decomposed and each MD acts as an agent to maximize the amount of the completed tasks (task throughput). 

\subsection{Multi-Agent Deep Deterministic Policy Gradient}
Although DDPG can adapt to the environment of multi-dimensional actions, it is difficult for a single super-agent to learn large-scale decentralized policies whose action space grows exponentially with the number of participants~\cite{jing2022multi}. MADDPG is an intuitive extension to the DDPG algorithm under a multi-agent system by decomposing a single monolithic agent into multiple simpler agents to reduce the dimensionality of the state and action spaces and thus overcome the scalability issue. In MADDPG, each agent makes the most suitable decision for itself, and multiple agents can achieve the common goal through cooperation. In this paper, we take advantage of MADDPG to train multiple agents for the optimization of multi-hop task offloading in vehicle-assisted MEC.


\subsection{State Space, Action Space and Reward Function}
1) \emph{State Space:} The state observed by MD $i$ at time slot $t$ is defined as
\begin{equation}
s_i(t)=\{s^w_i(t), \mathbf{s^s}(t), \mathbf{s^n}(t), \mathbf{s^l}(t)\},
\end{equation}
where $s^w_i(t)$ represents the number of tasks of MD $i$ arriving in time slot $t$, $\mathbf{s^s}(t)$ denotes the indexes of the selected ESs for all MDs at time slot $t$, $\mathbf{s^n}(t)$ denotes the number of MDs which select the same ES in time slot $t$, and $\mathbf{s^l}(t)$ denotes the remaining task size in the buffer at each ES in time slot $t$.

2) \emph{Action Space:} In the system, every MD has to decide the serving ES. Thus, the action of user $i$ at time slot $t$ is expressed as

\begin{equation}
a_i(t)=\{a^s_i(t)\},
\end{equation}
where $a^s_i(t)$ is the index of the ES selected by MD $i$. Besides, let $\mathcal{A}^s_i(t)$ denote the set of the selection actions by the feasible destinations. Therefore, action $a_i(t)$ is valid if $a^s_i(t) \in \mathcal{A}^s_i(t)$. Note that the routing path from MD $i$ to its destination ES will be uniquely determined if the ES is selected in MADDPG. For example, in this paper, we use the shortest path in terms of the travel distance between an MD and the associated ES.


3) \emph{Reward Function:} Since each MD intends to maximize its completed tasks while meeting the required e2e latency, the immediate reward is represented as
\begin{equation}
r_i(t)= c_i(t),
\end{equation}
where $c_i(t)$ is defined as the total size of accomplished tasks in time slot $t$ within the deadline, including tasks generated in the current time slot and those queued in the buffer. Note that the choice of the reward function is to approximately maximize the objective function defined in~\eqref{ob}, i.e., the number of tasks accomplished with latency requirements in the long run. 

The gained reward depends on the action of an MD, i.e., the MD gets an immediate reward $r_i(t)$ given observed state $s_i(t)$ and action $a_i(t)$ in time slot $t$. Each MD aims at learning the optimal policy which maximizes the long-term reward, which is given by
\begin{equation}
\mathcal{R}_i(t)=\max\; \mathbb{E}\left[\sum \limits_{k=0}^{T-1} \gamma^k r_i(k+t) \right],
\end{equation}
where $T$ is the number of consecutive time slots for calculating the long-term reward and $0$\textless $\gamma$ \textless $1$ is the discounting factor for determining the importance of the immediate reward and future rewards, where a smaller $\gamma$ means that more importance is given to the immediate reward.  

\subsection{The Training and Execution of MADDPG}
\begin{figure}[t]
\includegraphics[width=0.38\textwidth]{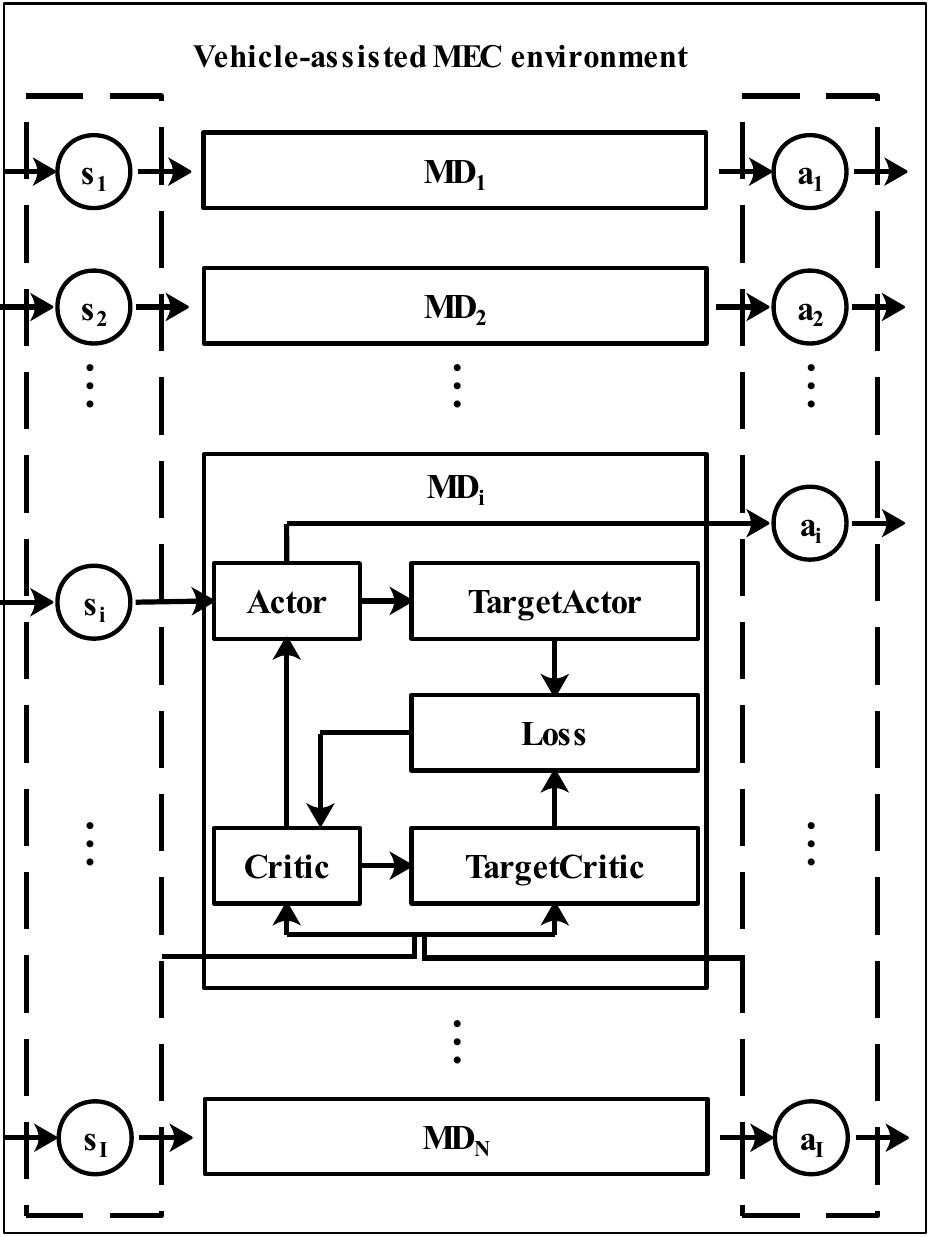}
\centering
\caption{The framework for an multi-agent deep deterministic policy gradient for task routing in vehicle-assisted MEC.}
\label{MADDPG}
 \end{figure}

Fig.\ref{MADDPG} illustrates the framework of MADDPG  with two main procedures: i) using the global information to train the \emph{critic network}, which is different from the traditional DDPG algorithm; and ii) using the local information to execute the \emph{actor network}. Suppose that there are $I$ agents (corresponding to the MDs in our system) in the vehicle-assisted MEC environment, in which we have two assumptions: i) the policy of each MD depends only on its own observed state, ii) the environment is unknown, and thus the reward for each agent and the next state after taking an action is unpredictable, which can only be acquired through the feedback from the environment.

\emph{Global training for the Critic network:} In the training of MADDPG, the \emph{actor network} selects an action according to the current state, and then the \emph{critic network} can calculate a Q value according to the state-action pair as feedback to the action. The \emph{critic network} is trained based on the estimated Q value and the actual Q value, and the \emph{actor network} updates the policy based on the feedback from the \emph{critic network}. To speed up the learning process of an MD, the input to the \emph{critic network} for training includes both its own observation and the observations (e.g., the states and actions) of other agents in the environment. The parameters in the \emph{critic network} are updated by minimizing the loss function based on Eq.~\eqref{loss_Critic}.


\emph{Local execution for the actor network:} When each MD is fully trained, each \emph{actor network} outputs appropriate actions according to its own state without the observed information from other MDs. The parameters in the \emph{actor network} are updated using gradient descent according to the loss function based on Eq.~\eqref{loss_Actor}. 

The training algorithm is summarized in Algorithm 1. We omit the introduction to Algorithm 1   due to the page limit. Please refer to~\cite{lowe2017multi} for more information.

\begin{algorithm}[]  
	\caption{Multi-Agent Deep Deterministic Policy Gradient for Task Routing in Vehicle-assisted MEC}
	\LinesNumbered 
	\KwIn{the number of arriving tasks for each MD, the locations of MDs, vehicles, and ESs, e2e latency requirements for MDs, computing capabilities of ESs, spectrum resources}
	\KwOut{task routing policy}
	\For{each episode}{
	Initialize a Gaussian noise $\mathcal{N}$ for action exploration \\
	Receive initial state $s$\\
		\For{each time slot $t=1,2,\cdot\cdot\cdot,T$}{
		For each agent $i$, select action $a_i=\boldsymbol{\pi}_{\theta_i}(s_i)+\mathcal{N}_t$ w.r.t the current policy and exploration\\
		Execute actions $a=(a_1,\cdot\cdot\cdot,a_I)$ and obtain rewards $r$ and new state $s^\prime$ from the environment\\
		Store $(\boldsymbol{s}, a, r,\boldsymbol{s}^{\prime })$ in Replay Buffer $\mathcal{D}$\\
		$s \leftarrow s^\prime$\\
		
			\For{each agent $i\in \mathcal{I}$}{
			Sample a random minibatch of $m$ samples $(\boldsymbol{s}^z, a^z, r^z,\boldsymbol{s}^{\prime z}), z=1,2,\cdot\cdot\cdot,m,$ from $\mathcal{D}$\\
			Set $y^z_i=r_i^z+\gamma Q_i^{\boldsymbol{\pi^\prime}}(\boldsymbol{s}^{\prime z},a_1^{\prime},\cdot\cdot\cdot, a_i\cdot\cdot\cdot,a_I^{\prime } )|_{a_i=\boldsymbol{ \pi^\prime}_i(s_i^z)}$\\
			$UpdateDDPG$
		}
		}
	}
\LinesNumbered
	Procedure: $UpdateDDPG$ \leavevmode\\
	Update \emph{critic network} by minimizing the loss function for $w_i$\leavevmode\\
			\begin{equation}\nonumber
           L(w_i)=\frac{1}{m}\sum \limits_{z=1}^m (y^z_i-Q(\phi(S^z),A^z,w_i))^2 
           \end{equation}
			Update \emph{actor network} by minimizing the loss function for $\theta_i$\leavevmode\\
        \begin{equation}\nonumber
        L(\theta_i)=-\frac{1}{m}\sum\limits_{z=1}^m Q(s^z_i,a^z_i,\theta_i)
        \end{equation}
        
		Update target network parameters for each MD $i$\leavevmode\\
		\begin{equation}\nonumber
		w_i^\prime \leftarrow \tau w_i+(1-\tau)w_i^\prime
      \end{equation}
      \begin{equation}\nonumber
      \theta_i^\prime \leftarrow \tau\theta_i+(1-\tau)\theta_i^\prime
        \end{equation}
\end{algorithm}

\section{Performance Evaluation}
We have conducted extensive studies to evaluate and compare the performance of the proposed MADDPG-based task routing in vehicle-assisted MEC with other benchmark solutions. The simulation experiments have been carried out on a ThinkPad X1 Carbon with a 4.7 GHz 12-Core Intel Core i7-1260P processor. The performance evaluation has been performed for two performance metrics: average throughput (task completion rate)  and success rate of the algorithms. 

\begin{figure}[t]
\includegraphics[width=0.6\textwidth]{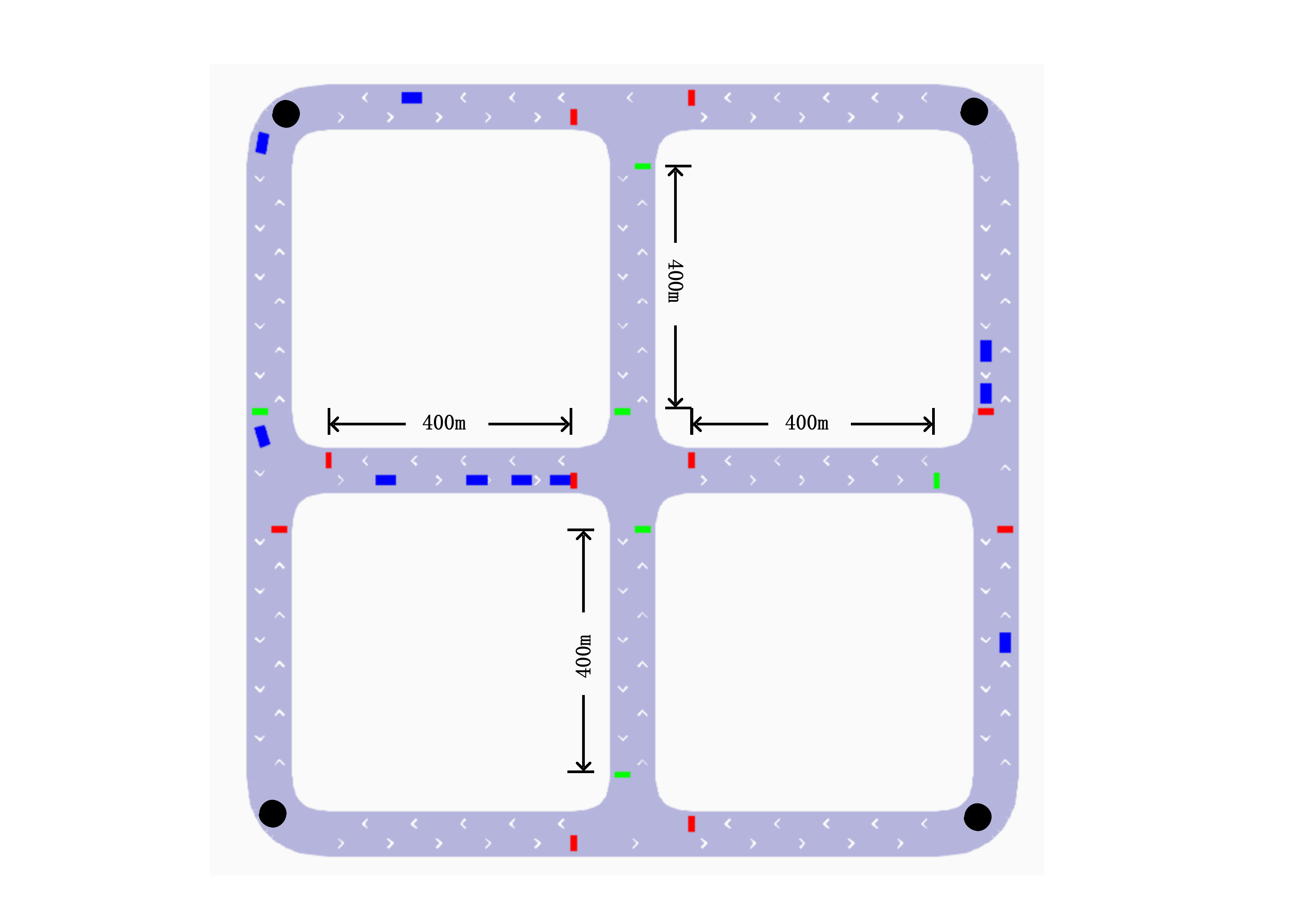}
\centering
\caption{The simulation scenario.}
\label{simulation}
 \end{figure}

\subsection{Simulation Settings}
1) \emph{Simulation Parameters:}  We consider a road network shown in Fig.~\ref{simulation}, where four ESs (i.e., black circles) are deployed as indicated in
the figure, and multiple vehicles (blue rectangles) are moving following traffic rules, i.e., subject to speed limits, safe distance, and traffic lights. The considered vehicle is 4 m in length and the safe distance between vehicles should be no less than 4 m. Besides, the speed limit in the considered road network is 60 km/h. The MDs' positions are randomly initialized in the road network at the beginning and fixed during the simulation. For each MD, the computing tasks are generated following a Poisson process and the task size is randomly distributed in the range of $[2,5]\times 10^5$ Kbits. In the simulation, we use $\beta$ to represent the probability of generating tasks for each MD in each time slot. Similar to~\cite{lei2020deep}, we consider a fair spectrum allocation rule among links in which the total bandwidth is proportionally allocated according to the size of transmitted tasks. Other simulation parameter settings are given in Table~\ref{tab:simulation}. We evaluate the performance within a duration of 100 seconds.

2) \emph{MADDPG Hyperparameters:} The actor network is a four-layer neural network with two fully connected hidden layers, each with 256 units and activated by sigmoid functions. The number of units of the input and the output layers are equal to the number of states and the number of actions, respectively. For the critic network, the input includes the actions produced by the actor network and the states. There are two hidden layers for the states and  one hidden layer for the actions before these two inputs are concatenated, after which there are two  fully connected hidden layers, each layer with 256 units and activated by ReLu functions. Finally, the critic network has an output layer to calculate the Q value for the given state-action pair, with no activation. Other hyperparameters
used for training MADDPG can be found
in Table~\ref{tab:hyperparameters}.
\begin{table}[t]
 \caption{\label{tab:simulation}Simulation Parameters}
 \centering
 \begin{tabular}{lcl}
  \toprule
  Parameter & Value \\
  \midrule
Coverage Radius of ESs&  200 m\\
Height of antenna&  1.5 m\\
Carrier frequency&  2800 MHz\\
Computation complexity& 1200 CPU cycles/bits\\
Bandwidth& $5$ MHz\\
Computing rate at ESs&$[1,2,3,4]\times 10^7$ cycles/s\\
Transmit power of MD& 1 W\\
Power of the Gaussian
noise& $5\times10^{-13}$ W\\
Discounting factor & 0.99\\
  \bottomrule
 \end{tabular}
\end{table}

\begin{table}[t]
 \caption{\label{tab:hyperparameters}Multi-Agent Deep Deterministic Policy Gradient Hyperparameters}
  \centering
 \begin{tabular}{lcl}
  \toprule
  Parameter & Value \\
  \midrule
 Replay buffer size & $10^5$ \\
 Minibatch size & 64 \\
 Gaussian noise, $\mathcal{N}(\mu, \sigma^2)$ & (0.15, $e^{-2}$) \\
 Learning rate of critic network &0.002  \\
 Learning rate of actor network &0.001  \\
 Update frequency of target network &0.005  \\
  \bottomrule
 \end{tabular}
\end{table}

\subsection{Compared Methods}
The proposed MADDPG-based task offloading method (referred to as \textit{MADDPG} in the following part) is compared with the following algorithms.

1) \textit{Single-hop:}  A direct task offloading from an MD to its ES via one-hop transmission is adopted considering both communication and computing resources, e.g., a game theoretic task offloading algorithm~\cite{deng2021throughput}. Once the workload exceeds the capacity of servers one-hop away, tasks will not be admitted into the system.


2) \textit{Multi-hop+Greedy:} A particular case of multi-hop task offloading with a naive solution. Once the workload exceeds the capabilities of local ESs, the remaining tasks will be delivered to ESs within the shortest distance via multi-hop transmission from the MD to the target ES.



We use two metrics to evaluate the effectiveness of the proposed MADDPG: the average service throughput (\textit{average throughput}) and the average success rate (\textit{success rate}). \textit{Average throughput} is the average size of completed tasks from MDs during $T$ time slots. \textit{Success rate} is the average ratio of the completed tasks (including the newly arrived and the buffered at ESs) to the generated tasks in each time slot. 

 \begin{figure*}[htbp]
\centering
\centering
\subfloat[Impact of task arrivals\\(when $I=20,N=4,J=4$).\label{throughput_task}]{\includegraphics[width=0.385\textwidth]{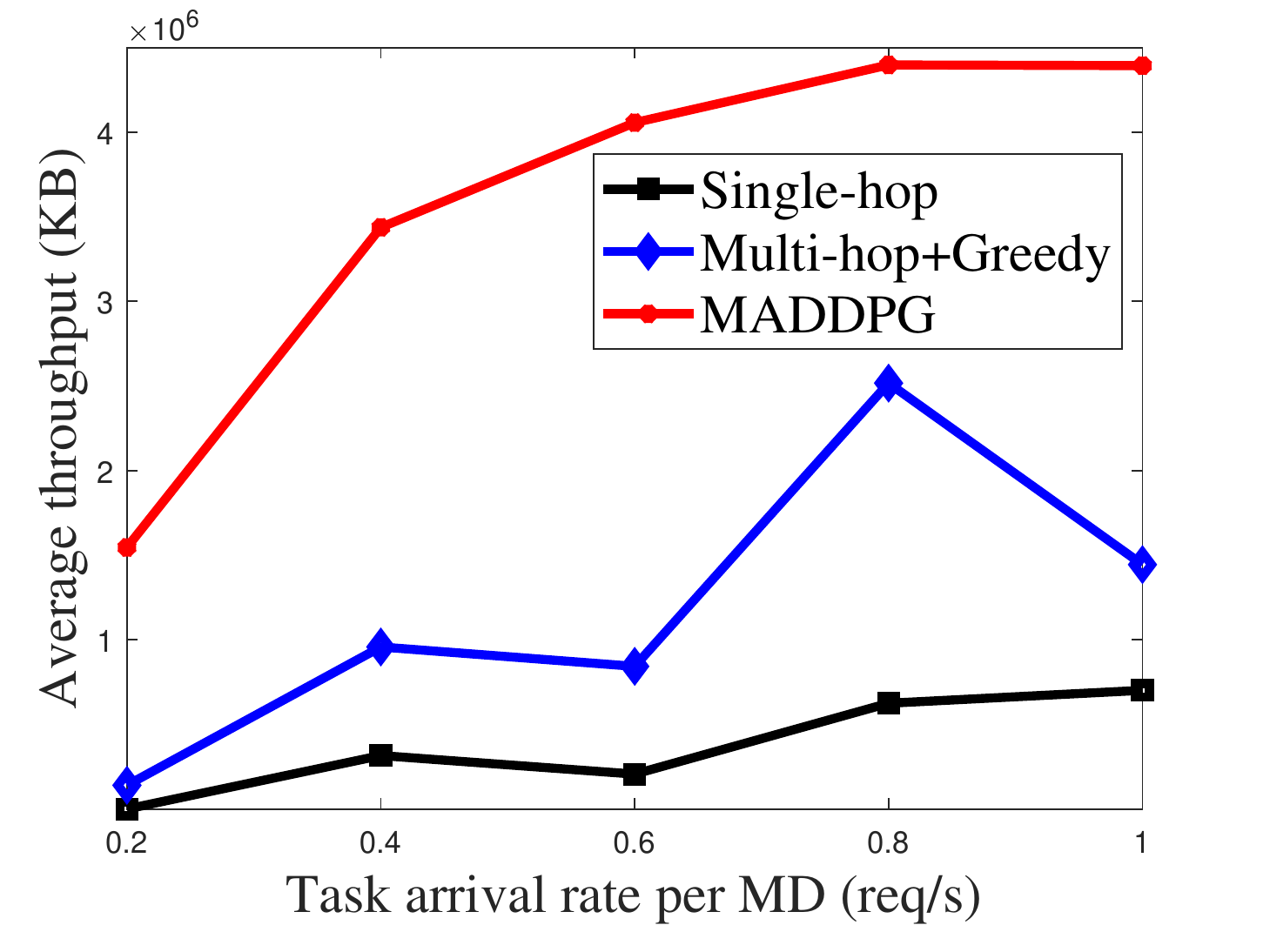}}
\subfloat[Impact of the number of MDs\\(when $\beta=0.5,N=4,J=4$).\label{throughput_user}]{\includegraphics[width=0.385\textwidth]{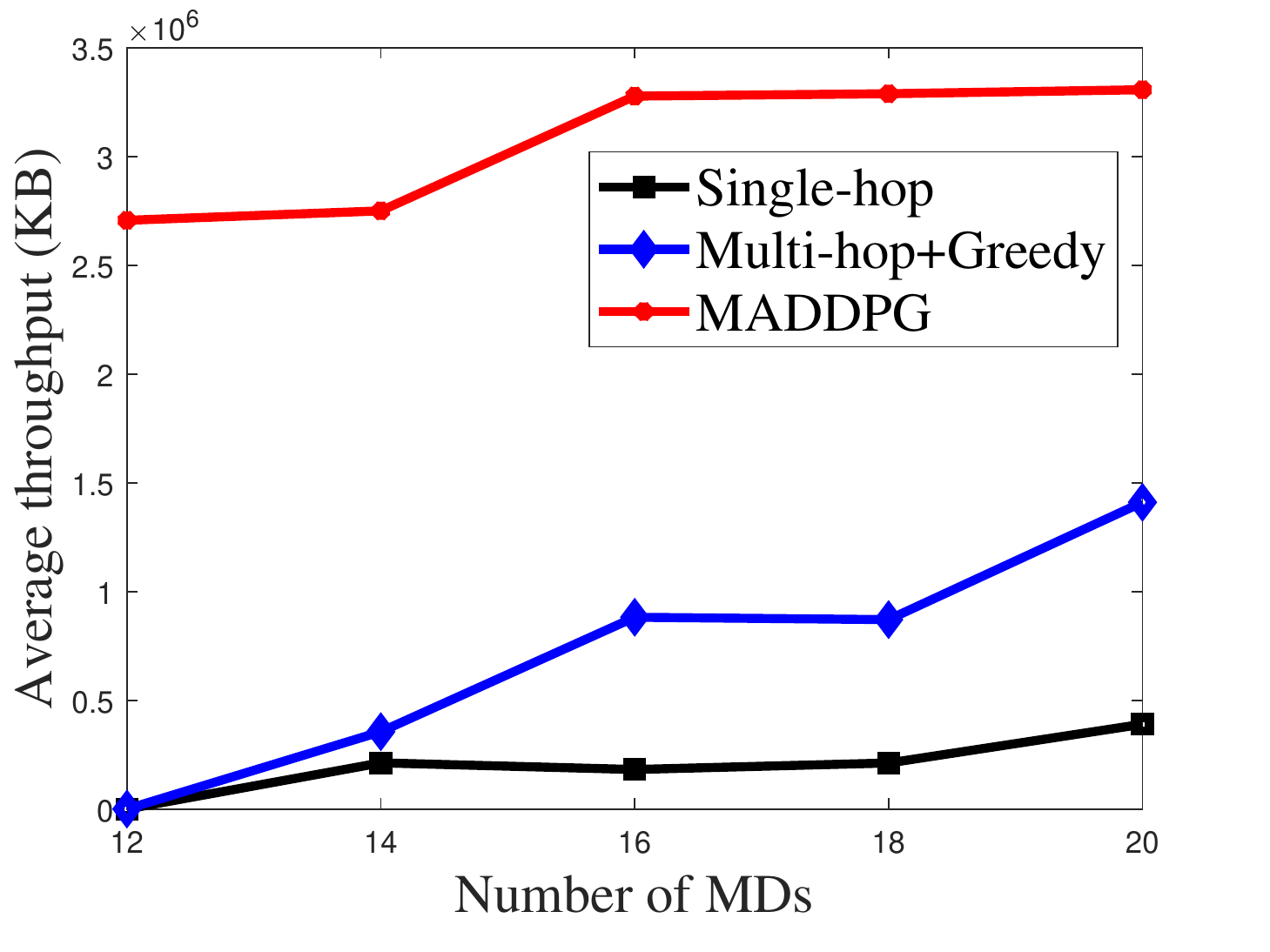}}
\hfill
\subfloat[Impact of the number of vehicles (when $\beta=0.5,I=20,J=2$).\label{throughput_vehicle}]{\includegraphics[width=0.385\textwidth]{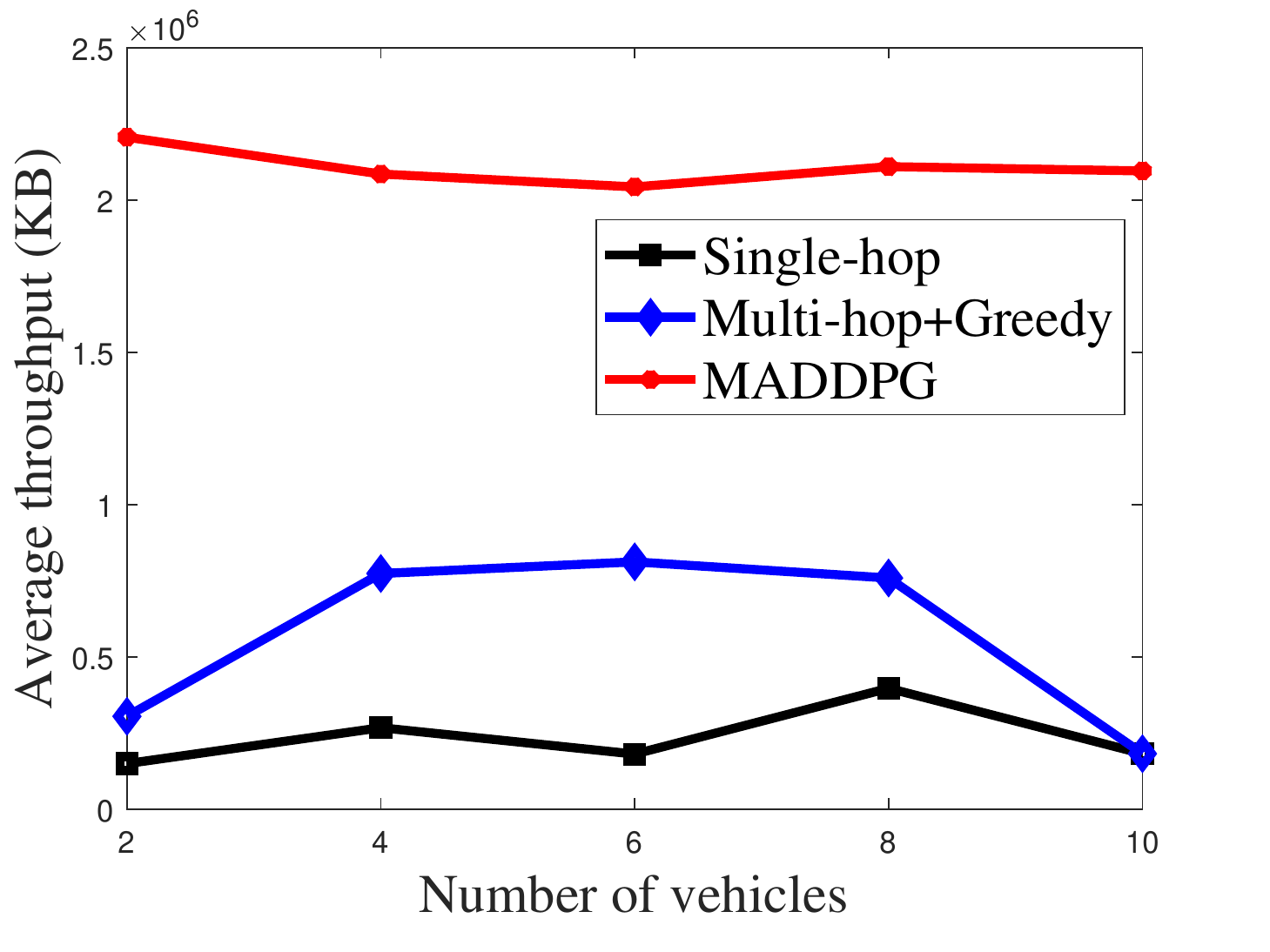}}
    \subfloat[Impact of the number of ESs \\(when $\beta=0.5,I=20,N=4$).\label{throughput_ES}]{\includegraphics[width=0.385\textwidth]{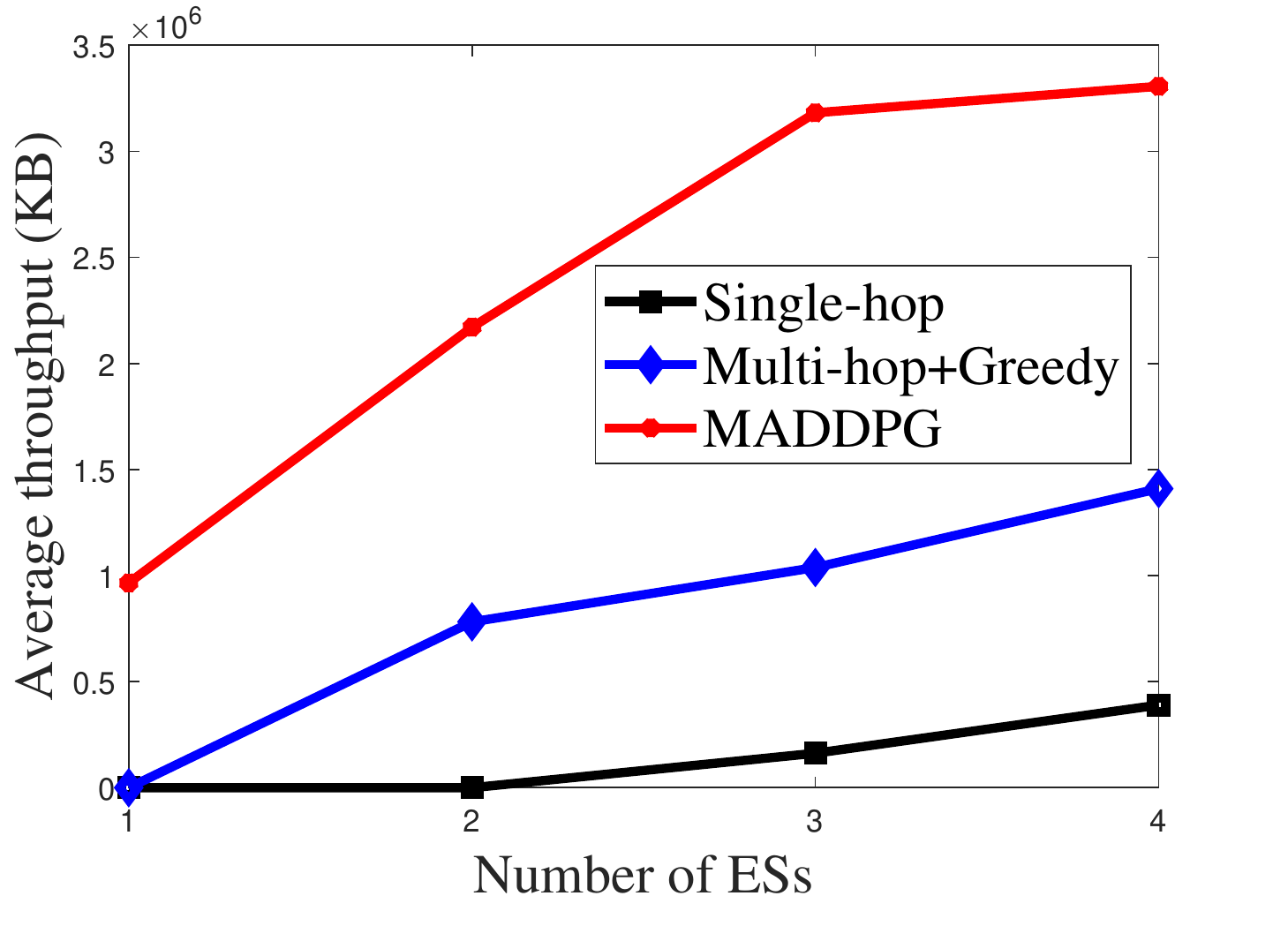}}
\caption{Evaluation of the average throughput\label{throughput}}
\end{figure*}

\subsection{Evaluation of the average throughput}
In Fig.~\ref{throughput}, we compare the average throughput of \textit{Single-hop}, \textit{Multi-hop+Greedy}, and \textit{MADDPG}. 
Fig.~\ref{throughput_task} demonstrates that the average throughput of \textit{MADDPG} climbs up as the task arrival rate increases. The reason is as follows. When the task arrivals from MDs or the number of ESs is large, the number of successfully completed tasks become large. Fig.~\ref{throughput_user} demonstrates that the average throughput of \textit{MADDPG} increases as the number of MDs increases when the number of MDs is in the range of $[10,20]$. Fig.~\ref{throughput_vehicle} shows that \textit{MADDPG} achieves a relatively stable average throughput when the number of vehicles varies from 2 to 10. The reason is that \textit{MADDPG} is capable of adaptively making multi-hop task offloading decisions by interacting with the environment. Fig.~\ref{throughput_ES} shows that the average throughput of \textit{MADDPG} increases with the number of ESs. This is because more ESs provides more resources for processing tasks in the system. Moreover, we observe that \textit{MADDPG} always has  much higher average throughput than~\textit{Single-hop} and~\textit{Multi-hop+Greedy}. That is because \textit{MADDPG} achieves load balancing via multi-hop transmissions between MDs and ESs, while \textit{Single-hop} can only exploit ESs one-hop away and \textit{Multi-hop+Greedy} may cause computing overload and network congestion by blindly selecting ESs.

\subsection{Evaluation of the success rate}
\begin{figure*}[htbp]
\centering
\centering
\subfloat[Impact of task arrivals\\(when $I=20,N=4,J=4$).\label{SuccessRate_task}]{\includegraphics[width=0.385\textwidth]{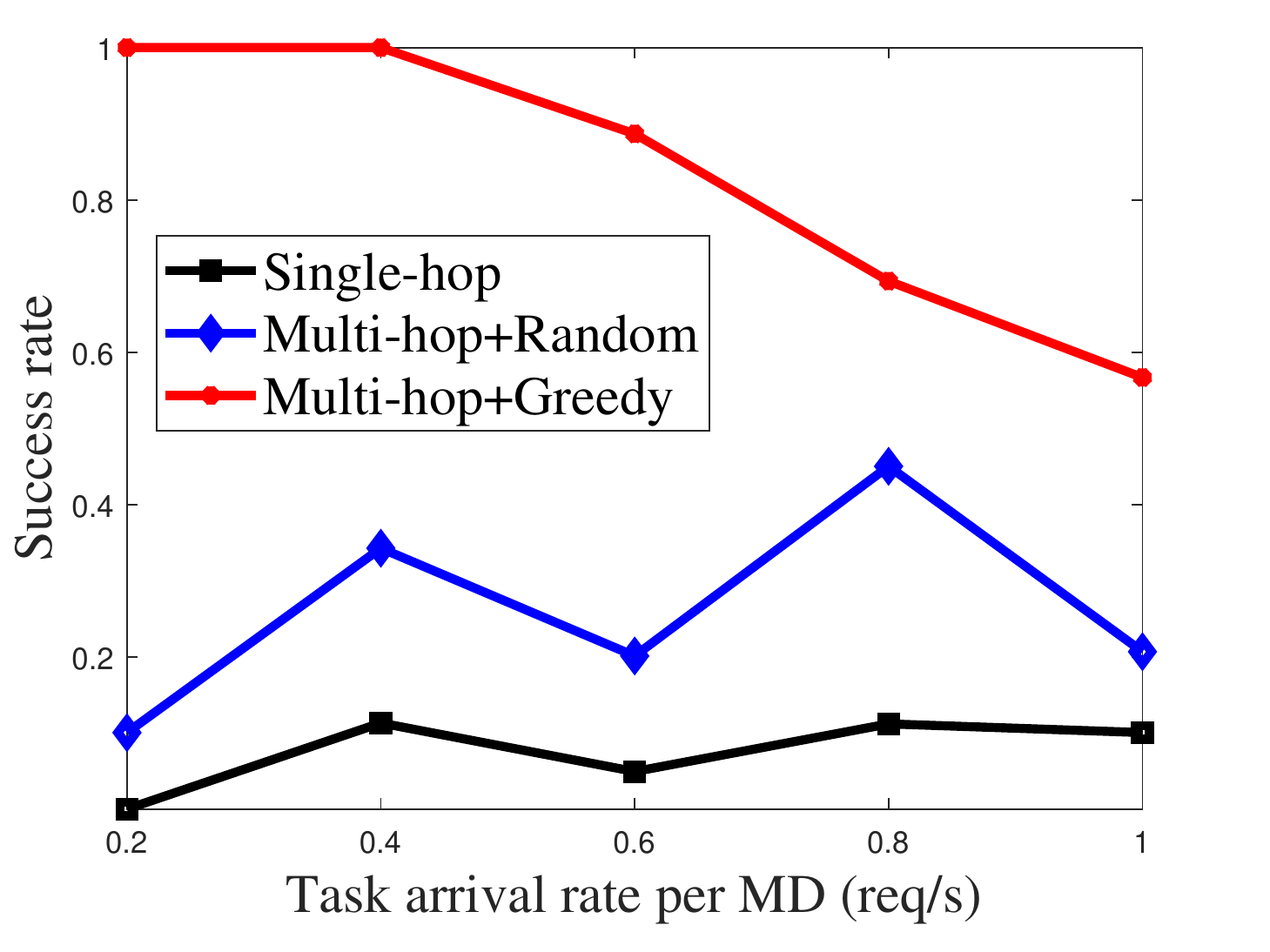}}
    \centering
    \subfloat[Impact of the number of MDs\\(when $\beta=0.5,N=4,J=4$).\label{SuccessRate_user}]{\includegraphics[width=0.385\textwidth]{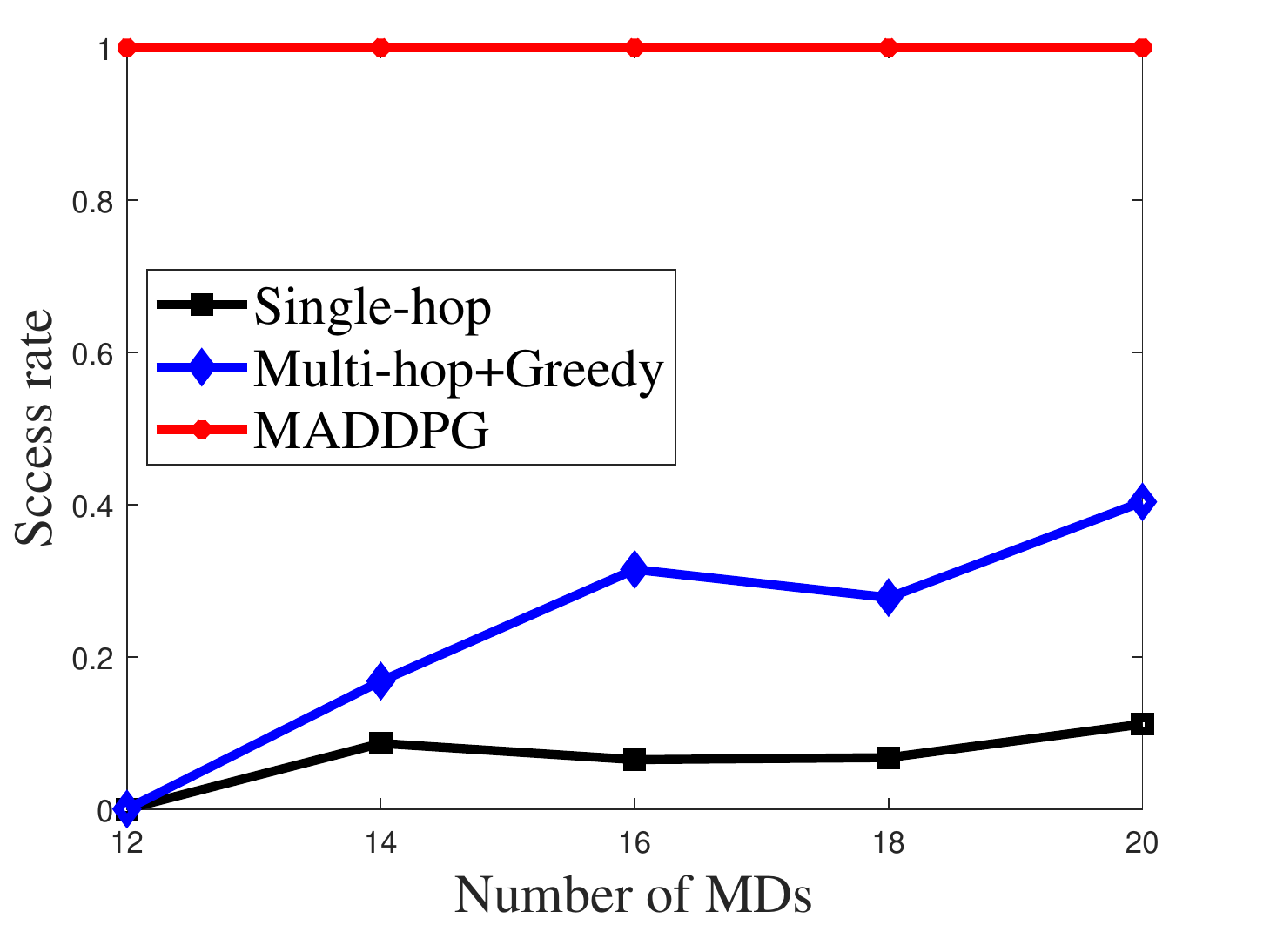}}
\hfill
\subfloat[Impact of the number of vehicles (when $\beta=0.5,I=20,J=2$).\label{SuccessRate_vehicle}]{\includegraphics[width=0.385\textwidth]{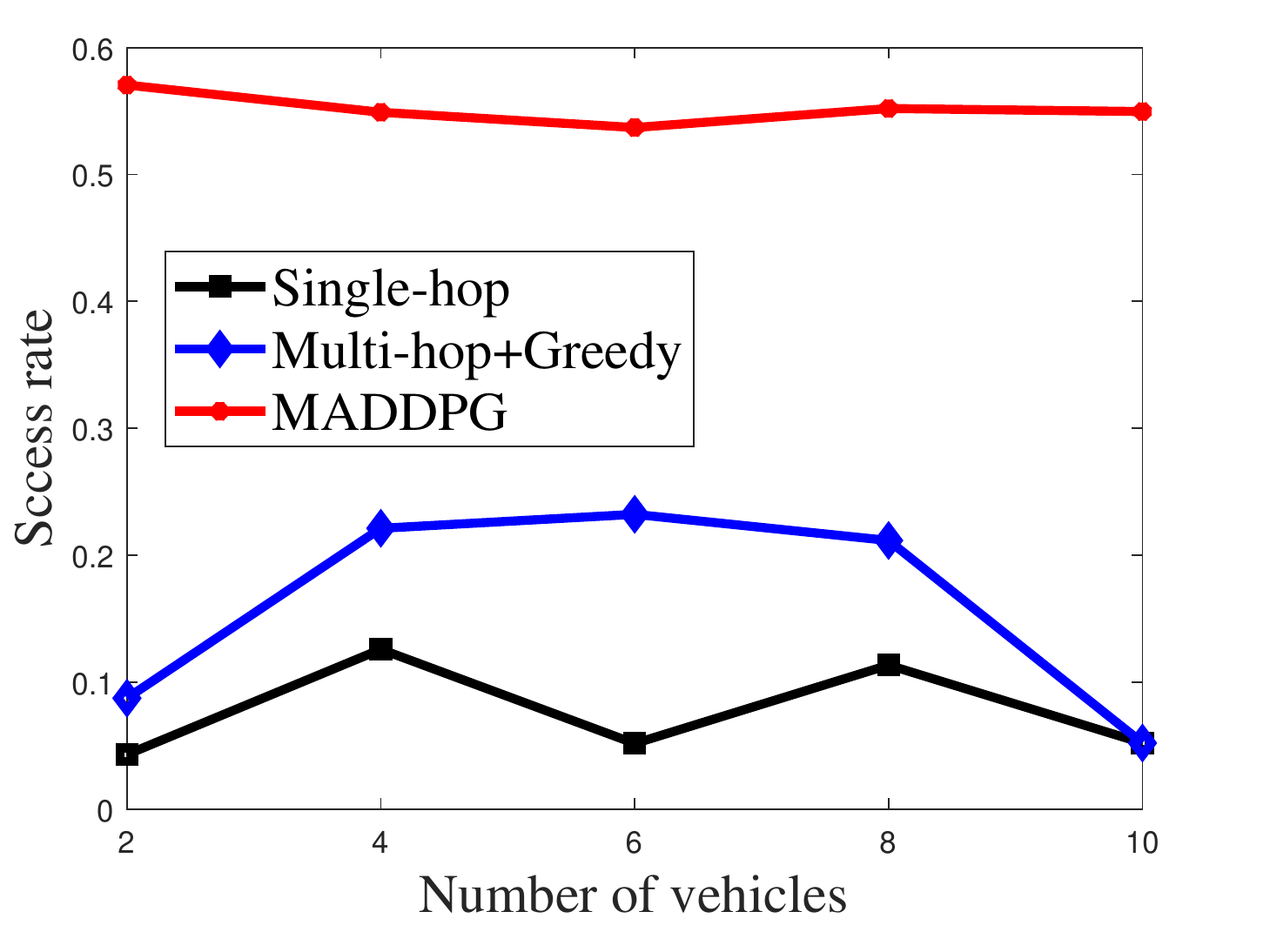}}
\subfloat[Impact of the number of ESs \\(when $\beta=0.5,I=20,N=4$).\label{SuccessRate_ES}]{\includegraphics[width=0.385\textwidth]{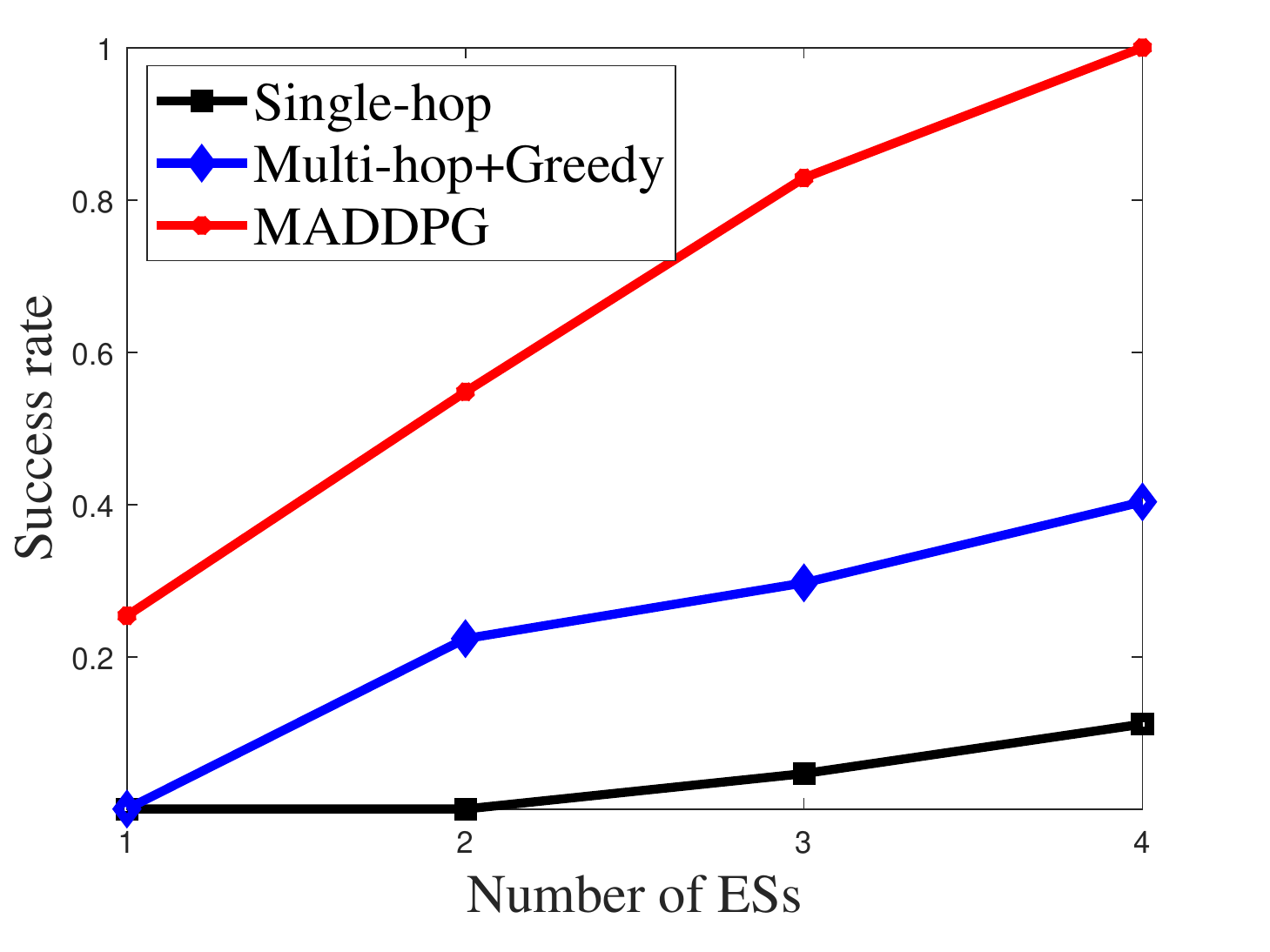}}
    \caption{Evaluation of the success rate\label{SuccessRate}}
\end{figure*}

Fig.~\ref{SuccessRate} shows the success rate comparison between \textit{MADDPG}
with \textit{Single-hop} and \textit{Multi-hop+Greedy}. 
Fig.~\ref{SuccessRate_task} shows the success rate of three task
offloading mechanisms against the task arrivals. We observe that tasks generated by MDs can be almost completed by \textit{MADDPG} when the task arrival rate is small. In addition, \textit{Single-hop} and \textit{Multi-hop+Greedy} have unstable performance due to highly dynamic network environments, e.g., randomly distributed MDs and frequently moving vehicles. In Fig.~\ref{SuccessRate_user}, we observe that tasks can be almost accomplished when the number of MDs varies from 10 to 20. Fig.~\ref{SuccessRate_vehicle} demonstrates that the success rate of three task
offloading mechanisms against the number of vehicles. Fig.~\ref{SuccessRate_ES} shows the success rate of three task
offloading mechanisms against the number of ESs. We can see that the success rate of \textit{MADDPG} is approaching 1 as the number of ESs increases while \textit{Single-hop} and \textit{Multi-hop+Greedy} have much lower success rate. This is because \textit{MADDPG} enables MDs to adaptively select destination ESs while \textit{Single-hop} can only utilize local ES resources. Besides, \textit{Multi-hop+Greedy} may incur selection conflicts of destination ESs, which causes performance degradation to some extent. Another observation is that \textit{MADDPG} keeps the highest success rate among the three schemes given any task arrival rate, the number of MDs, the number of vehicles, and the number of ESs, respectively. 

\section{Conclusion}
By allowing edge servers multi-hop away to share the computing workload, the multi-hop MEC enables more edge servers to share their computing resources. In this paper, we have proposed such a novel multi-hop task offloading approach for MEC systems with the assistance of vehicles to enhance the system capacity while satisfying users’ e2e latency requirements. In a highly dynamic and complicated system, we have taken into account several practical factors, such as vehicular mobility, spectrum availability, and computing capabilities, to obtain the association between users and edge servers potentially multi-hop away, a scenario rarely considered before. By employing multi-agent reinforcement learning, each end user acts as an agent to efficiently and adaptively make offloading policy achieve high aggregated throughput subject to its end-to-end latency requirement and resource limitations in a time-varying vehicular network. Extensive simulations have demonstrated that the proposed MADDPG-based task offloading scheme can increase the number of completed tasks while providing latency guarantee through adaptive load balancing among edge servers possibly multi-hop away achieved by running effective reinforcement learning mechanisms. 


\bibliographystyle{IEEEtran}
\bibliography{reference}

\begin{IEEEbiography}[{\includegraphics[width=1in,height=1.25in,clip,keepaspectratio]{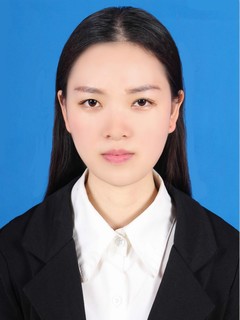}}]{Yiqin Deng} is currently a Postdoctoral Research Fellow with the School of Control Science and Engineering, Shandong University. She received her B.S. degree in project management from Hunan Institute of Engineering, Xiangtan, China, in 2014, and her M.S. degree in software engineering and her Ph.D. degree in computer science and technology from Central South University, Changsha, China, in 2017 and 2022, respectively. She was a visiting researcher at the University of Florida, Gainesville, from 2019 to 2021. Her research interests include Edge/Fog computing, Internet of Vehicles, and Resource Management.
\end{IEEEbiography}
\begin{IEEEbiography}[{\includegraphics[width=1in,height=1.25in,clip,keepaspectratio]{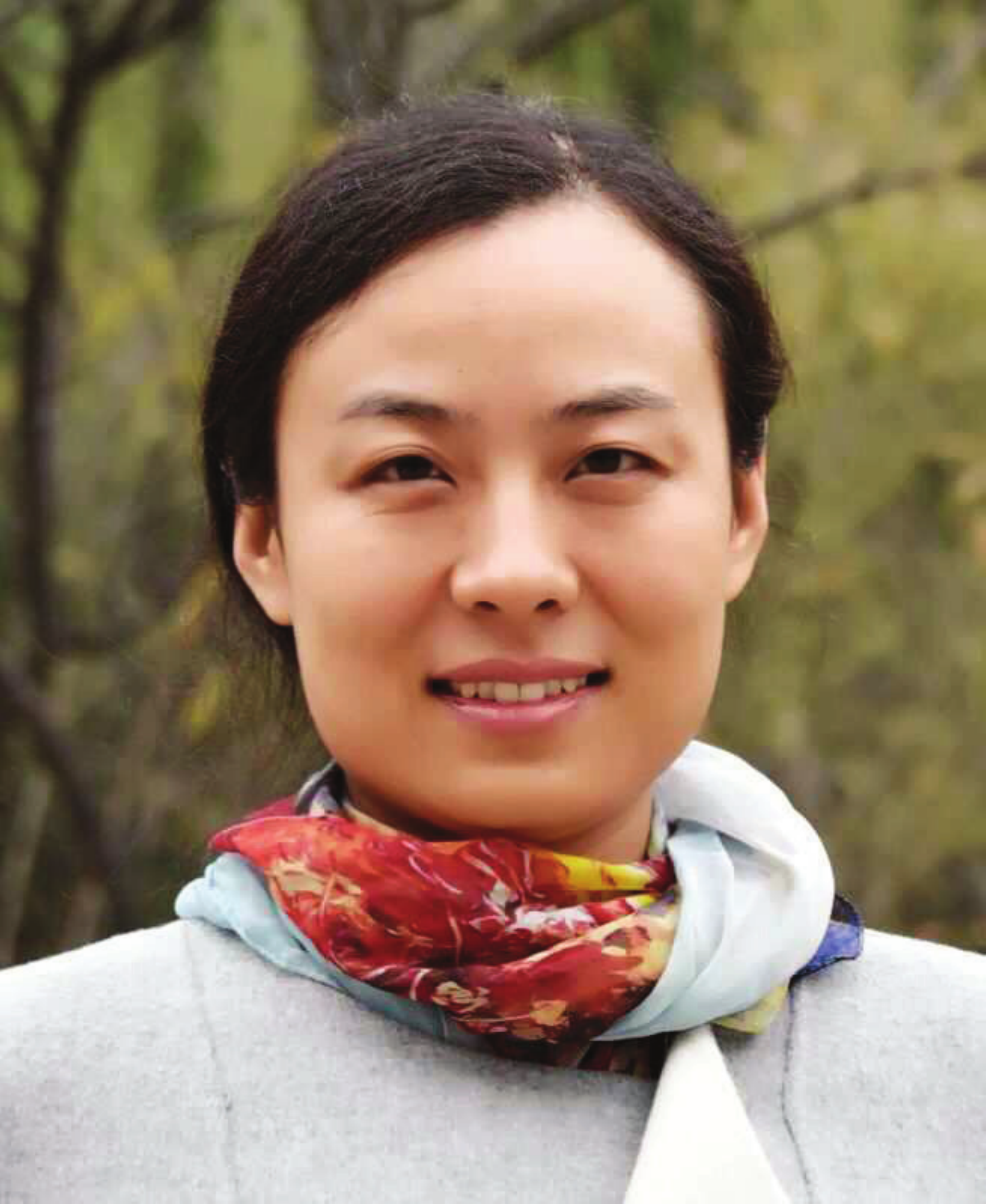}}]{Haixia Zhang}(M'08-SM'11) received the B.E. degree from the Department of Communication and Information Engineering, Guilin University of Electronic Technology, China, in 2001, and the M.Eng. and Ph.D. degrees in communication and information systems from the School of Information Science and Engineering, Shandong University, China, in 2004 and 2008, respectively. From 2006 to 2008, she was with the Institute for Circuit and Signal Processing, Munich University of Technology, as an Academic Assistant. From 2016 to 2017, she was a Visiting Professor with the University of Florida, USA. She is currently a Distinguished Professor with Shandong University. Her current research interests include industrial Internet of Things (IIoT), resource management, mobile edge computing, and smart communication technologies. Dr. Zhang serves on editorial boards of the IEEE Transactions on Wireless Communications, IEEE Wireless Communication Letters, and China Communications. She has been serving as TPC member, session chair, invited speaker and keynote speaker for conferences. 
\end{IEEEbiography}
\begin{IEEEbiography}[{\includegraphics[width=1in,height=1.25in,clip,keepaspectratio]{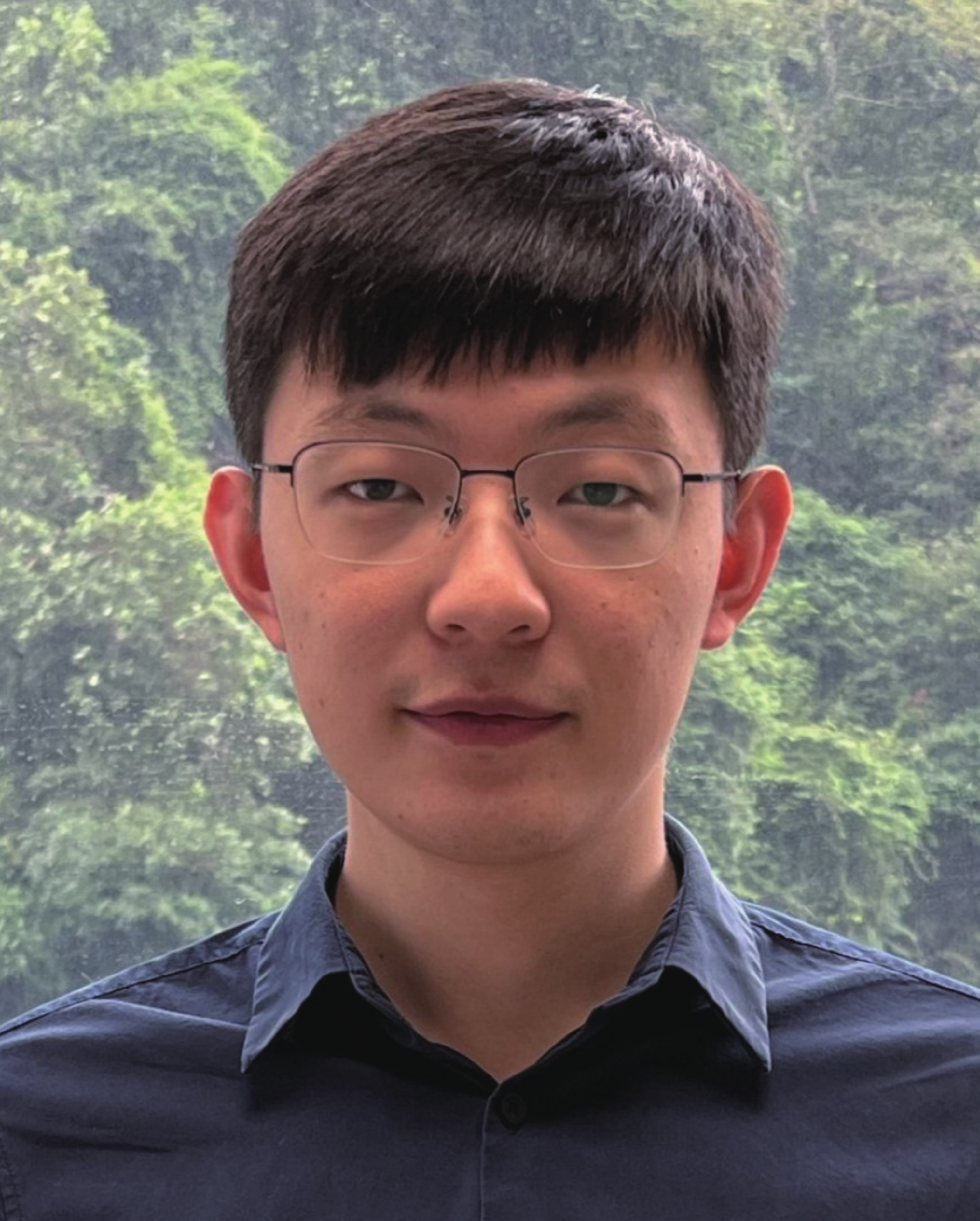}}]{Xianhao Chen}
 is currently an assistant professor with the Department of Electrical and Electronic Engineering, the University of Hong Kong. He obtained the Ph.D. degree from the University of Florida in 2022, and received the B.Eng. degree from Southwest Jiaotong University in 2017. His research interests include wireless networking and machine learning.
\end{IEEEbiography}
\begin{IEEEbiography}[{\includegraphics[width=1.2in,height=1.3in,clip,keepaspectratio]{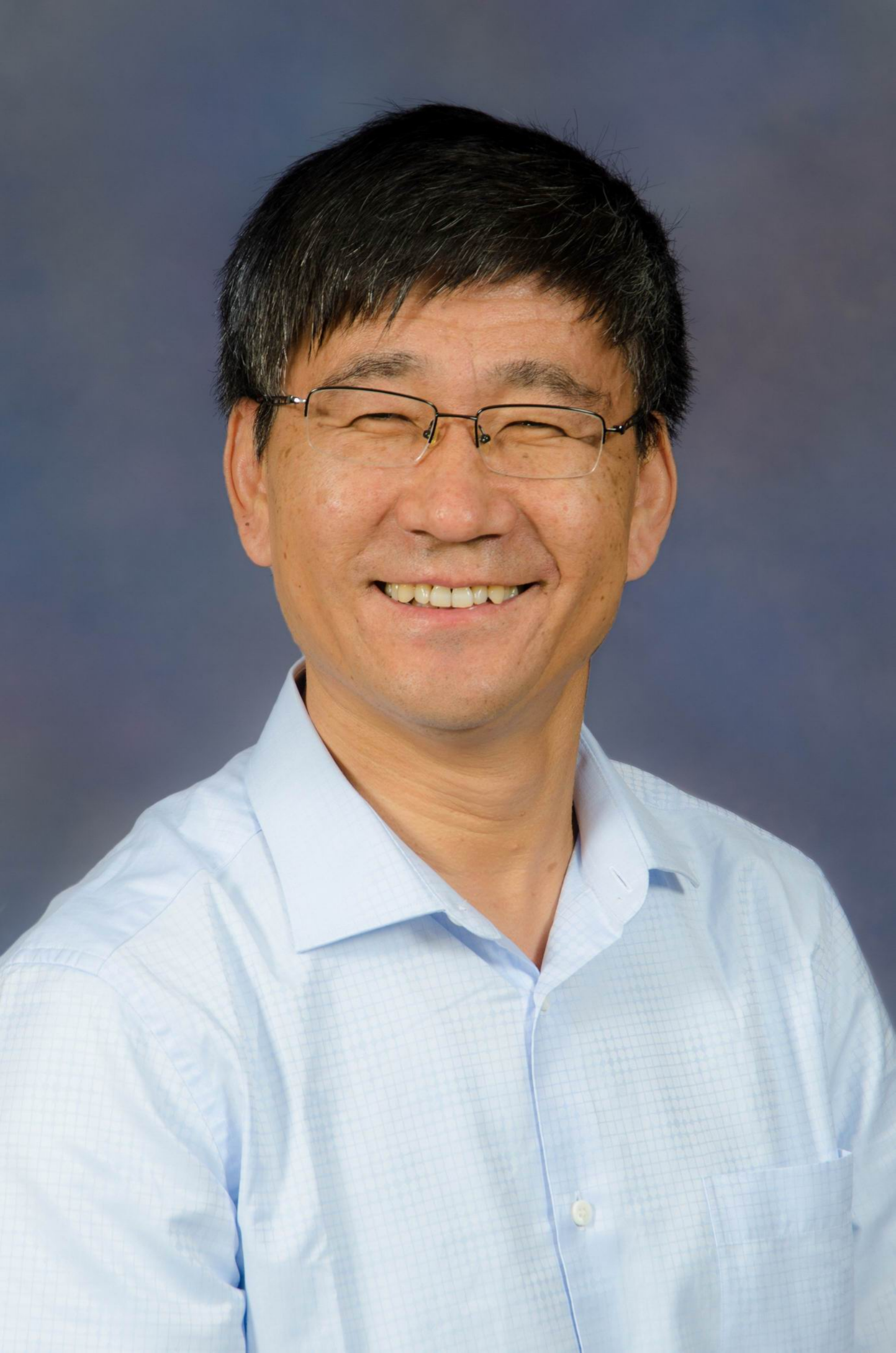}}]{Yuguang Fang} (S’92, M’97, SM’99, F’08) received an MS degree from Qufu Normal University, China in 1987, a PhD degree from Case Western Reserve University in 1994, and a PhD degree from Boston University in 1997. He joined the Department of Electrical and Computer Engineering at University of Florida in 2000 as an assistant professor, then was promoted to associate professor in 2003, full professor in 2005, and distinguished professor in 2019, respectively. Since 2022, he has been the Chair Professor of Internet of Things with Department of Computer Science at City University of Hong Kong.

Dr. Fang received many awards including the US NSF CAREER Award, US ONR Young Investigator Award, 2018 IEEE Vehicular Technology Outstanding Service Award, IEEE Communications Society AHSN Technical Achievement Award (2019), CISTC Technical Recognition Award (2015), and WTC Recognition Award (2014), the Best Paper Award from IEEE ICNP (2006), and 2010-2011 UF Doctoral Dissertation Advisor/Mentoring Award. He was the Editor-in-Chief of IEEE Transactions on Vehicular Technology (2013-2017) and IEEE Wireless Communications (2009-2012) and has served on several editorial boards of premier journals. He also served as the Technical Program Co-Chair of IEEE INFOCOM’2014. He is a fellow of ACM, IEEE, and AAAS.


\end{IEEEbiography}

\end{document}